\documentclass[aps,twocolumn,showpacs]{revtex4}
\usepackage{graphicx}
\usepackage{epsfig}

\newcommand{\beq}{\begin{equation}}
\newcommand{\eeq}{\end{equation}}
\newcommand{\beqr}{\begin{eqnarray}}
\newcommand{\eeqr}{\end{eqnarray}}

\begin{document}

\title{Spin-Squeezing and Light Entanglement in Coherent Population Trapping}

\author{A. Dantan$^1$, J. Cviklinski$^2$, E. Giacobino$^2$, M. Pinard$^2$}

\affiliation{$^1$ Laboratoire Charles Fabry de l'Institut d'Optique,
F91403 Orsay cedex, France\\$^{2}$ Laboratoire Kastler Brossel,
UPMC, 4 place Jussieu, F75252 Paris cedex 05, France}

\begin{abstract}
We show that high squeezing and entanglement can be generated at the
output of a cavity containing atoms interacting with two fields in a
Coherent Population Trapping situation, on account of a non-linear
Faraday effect experienced by the fields close to a dark-state
resonance in a cavity. Moreover, the cavity provides a feedback
mechanism allowing to reduce the quantum fluctuations of the ground
state spin, resulting in strong steady state spin-squeezing.
\end{abstract}


\maketitle

Coherence-mediated effects in atomic media have received
considerable attention in connection with magnetometry
\cite{budker02}, coherent population trapping \cite{arimondo},
electromagnetically induced transparency \cite{fleischhauer05},
slow-light \cite{budker99}, non-linear optics \cite{nl} or atomic
spin-squeezing \cite{dantanpra03}. When two fields of about the same
strength interact resonantly with three-level $\Lambda$-type atoms -
situation commonly referred to as Coherent Population Trapping (CPT)
- the atoms are pumped into a superposition of the ground state
levels, which is a state with maximum coherence. Close to the CPT
(or dark-) resonance the fields experience a strong dispersive
effect, but little absorption \cite{fleischhauer05}. Consequently,
several schemes have been studied which take advantage of this
strong non-linearity to generate quantum correlations and squeezing
in the fields under EIT or CPT conditions
\cite{fleischhauer,qnd,garrido}.

Following the recent squeezed and entangled light states generation
with cold atoms \cite{josseprl03,josseprl04}, we study in this
Letter the interaction of atoms inside an optical cavity with two
field modes close to a dark-resonance. We first show that a
multistable behavior for the intracavity light may occur close to
the CPT resonance on account of a non-linear polarization
self-rotation effect experienced by the light \cite{sr,ps}. Such
polarization instabilities have been observed with cold atoms
\cite{josseprl03} and thermal vapor cells \cite{ps}. We then
calculate the field noise spectra and predict that strong
correlations exist between the fields exiting the cavity and that
squeezing and entanglement can be generated close to the switching
threshold and for a wide range of parameters. In contrast with the
experiments of Refs.~\cite{josseprl03,josseprl04,hsu} the squeezing
and the entanglement are not deteriorated by excess atomic noise due
to optical pumping processes and could be efficiently generated
either with cold atoms or thermal vapor cells. Last, we show that
under appropriate conditions on the cavity induced feedback the
ground state atomic spin fluctuations may also be strongly squeezed,
and subsequently read out using techniques developed within the
context of quantum memory \cite{dantanpra04}.

We consider $N$ $\Lambda$-like atoms with ground-states 1 and 2,
interacting with two modes of the field, $A_1$ and $A_2$. To
simplify the discussion we turn to the symmetrical case of incident
fields with equal power and will choose the parameters so that the
Rabi frequencies of both transitions are equal: $\Omega_i=\Omega$
($i=1,2$) and close to one-photon resonance. This situation
corresponds to Coherent Population Trapping, since the atoms are
pumped into a superposition of levels 1 and 2 - the so-called
\textit{dark-state} - which is decoupled from the fields
\cite{arimondo}. For $\Omega_1=\Omega_2$, this dark-state
corresponds to a state with maximum ground state coherence:
$|D\rangle=(|1\rangle-|2\rangle)/\sqrt{2}$. Although this is not
essential, let us note that, if levels 1 and 2 are Zeeman sublevels,
the incident field can then be considered as being linearly
polarized. As we will see later this facilitates the physical
discussion and provides a simple picture in terms of the Stokes
polarization vector and the collective spin formalism.

\begin{figure}
  \includegraphics[width=7cm]{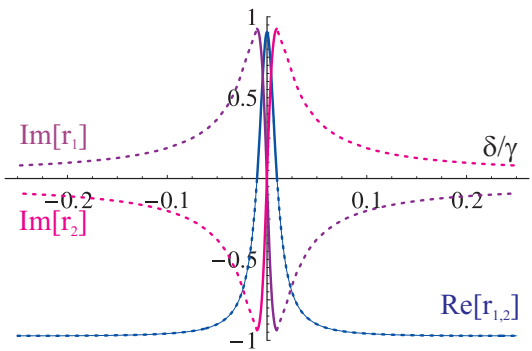}
  \caption{Real and imaginary parts of the reflectivity coefficients of the cavity when the two-photon detuning $\delta$ is varied.
  Parameters: $C=100$, $I=1$, $\varphi=0$. The dashed parts corresponds to unstable solutions for the linearly-polarized light.}\label{fig:cpt}
\end{figure}

If the fields are symmetrically detuned with respect to the atomic
resonance: $\Delta_2=-\Delta_1=\delta$, and for opposite cavity
detunings $\Delta_{c1}=-\Delta_{c2}=\kappa\varphi$, the intracavity
intensities are indeed symmetrical with respect to each mode and
satisfy \beqr\label{eq:I} I^{in}&=&I\left[\left(1+A\right)^2+
\left(\varphi-\phi_{nl}\right)^2\right]\\\label{eq:A}
A&=&C\frac{\bar{\delta}^2}{I^2+\bar{\delta}^2+I\bar{\delta}^2+\bar{\delta}^4}\\\label{eq:phinl}
\phi_{nl}&=&C\frac{\bar{\delta}(I-\bar{\delta}^2)}{I^2+\bar{\delta}^2+I\bar{\delta}^2+\bar{\delta}^4}\eeqr
with $I=\Omega^2/\gamma^2$, $\bar{\delta}=\delta/\gamma$, $\gamma$
the optical dipole decay rate, $\kappa$ the cavity bandwidth and
$C=g^2N/T\gamma$ the usual cooperativity parameter, $g$ being the
atom-field coupling constant and $T$ the intensity transmission of
the coupling mirror. In the vicinity of the CPT resonance
($\delta=0$), levels 1 and 2 are equally populated, and the ground
state coherence is real and maximal: $\langle J_x\rangle\simeq
-N/2$. As illustrated in Fig.~\ref{fig:cpt}, the absorption of the
fields is drastically reduced within a narrow transparency window,
while there is a strong change in the dispersion.

Fully on resonance the medium is rendered transparent in steady
state for both fields, so that, if the input fields are in a
coherent state, the output fields will also be uncorrelated.
However, in the vicinity of such a dark resonance, Eqs.
(\ref{eq:A}-\ref{eq:phinl}) show that the non-linearity and
absorption $A$ of the medium are given by
\begin{equation} A\sim \frac{C\delta^2\gamma^2}{\Omega^4},
\hspace{1cm}\phi_{nl}\sim\frac{C\delta\gamma}{\Omega^2}\end{equation}
when $\Omega\gg\gamma,\delta$. These quantities are to be compared
with the absorption and Kerr non-linearity in a two-level system
\cite{lambrecht96,josseprl03}
\begin{equation}
A\sim\frac{C\gamma^2}{2\Delta^2},\hspace{1cm}\phi_{nl}\sim\frac{C\gamma\Omega^2}{\Delta^3}\end{equation}
($\Delta\gg\gamma,\Omega$). For squeezing or entanglement generation
the figure of merit of both systems can be evaluated qualitatively
by comparing the non-linearity to absorption ratio $\phi_{nl}/A$. In
both cases these ratios take on a similar form:
$(\phi_{nl}/A)_{CPT}=\Omega^2/\delta\gamma$ and
$(\phi_{nl}/A)_{Kerr}=2\Omega^2/\Delta\gamma$, but, the sharpness of
the dark resonance allows to reach a strong non-linearity as well as
little absorption with much lower intensities \cite{nl}.\\

\begin{figure}
  \includegraphics[width=8.5cm]{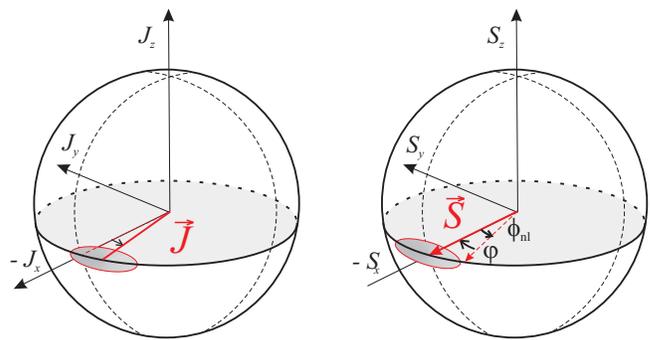}\\
  \caption{Self-rotation induced on the field close to the CPT resonance: a non-zero detuning $\delta$
  causes the spin to rotate in the equatorial plane of the Bloch sphere by an angle proportional to $\delta$.
  Because of the Faraday effect, the Stokes vector also tends to rotate by an angle $\phi_{nl}$ proportional to
  $\delta$ in the equatorial plane of the Poincar\'{e} sphere, but the detuned cavity brings it back along $x$.
  The ellipsoids represent the quantum fluctuations of the Bloch and Stokes vectors.}\label{fig:SRpolarisation}
\end{figure}

A first consequence of this non-linearity is the appearance of a
multistable behavior for the intracavity light \cite{sr}. As shown
in Fig.~\ref{fig:cpt} the symmetrical solution - always stable for
$\delta=0$ - is unstable above a certain threshold when the
two-photon resonance condition is no longer ensured. From
Eq.~(\ref{eq:I}) the stability condition corresponds to
\begin{equation}\delta\leq \delta_s=\sqrt{1+\varphi^2}\frac{\Omega^2}{\gamma C}\hspace{0.7cm}(C\gg 1, I\gg \bar{\delta}).\label{eq:seuil}\end{equation}
If one considers the $A_1$ and $A_2$ modes to be
circularly-polarized, this threshold can be traced to a
coherence-induced non-linear \textit{self-rotation} \cite{sr,ps}.
For the chosen Rabi frequencies the incident field is
linearly-polarized along the $y$-axis, which means that the Stokes
vector is aligned along $-Ox$ in the Poincar\'{e} sphere
(Fig.~\ref{fig:SRpolarisation}). On resonance ($\delta=0$) the
ground state spin in the Bloch sphere is parallel to the Stokes
vector : $\langle J_x\rangle=-N/2$, $\langle J_y\rangle=\langle
J_z\rangle=0$. If one lifts the ground state sublevel degeneracy
(with a longitudinal magnetic field for instance), the intracavity
fields experience opposite non-linear phase-shifts as can be seen
from Fig.~\ref{fig:cpt}. The Stokes vector would thus tend to rotate
in the equatorial plane of the Poincar\'{e} sphere under the
influence of this non-linear Faraday effect
(Fig.~\ref{fig:SRpolarisation}), but, if the phase-shift stays
smaller than the cavity and atomic losses then the Stokes vector
stays along the $x$-axis. Because of the Faraday effect the ground
state spin rotates in the equatorial plane by a small angle
proportional to $\delta$ at first order: \beqr \langle
J_x\rangle\simeq -\frac{N}{2},\hspace{0.3cm}\langle J_y\rangle\simeq
\frac{N}{2}\frac{\delta\gamma}{\Omega^2}.\eeqr

\begin{figure}
  \includegraphics[width=7.5cm]{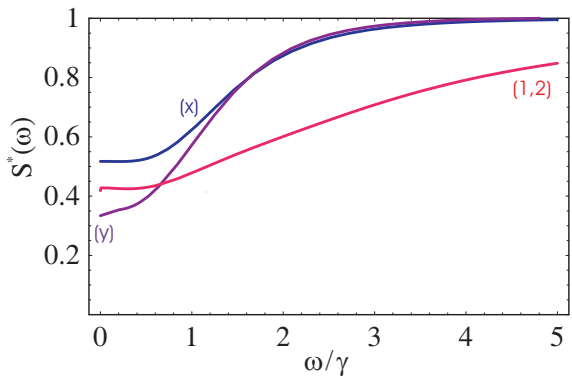}\\
  \caption{Squeezing spectra of the circularly- and linearly-polarized modes, $A_{1,2}$ and $A_{x,y}$ [as defined by Eq. (\ref{eq:AxAy})].
  Parameters: $C=100$, $\kappa=2\gamma$
  $\bar{\delta}=1$, $\varphi=1$, $I=144$.}\label{fig:squeezingdelta2}
\end{figure}

We now turn to the modification of the outgoing field noise spectra
in the vicinity of the CPT resonance. We calculate these spectra in
a standard fashion by linearizing the equations around the
semi-classical state corresponding to a working point in the stable
range defined previously. The noise spectrum
$S_{X_{\theta}}(\omega)$ of any quadrature
$X_{\theta}=Ae^{-i\theta}+A^{\dagger}e^{i\theta}$ can be obtained
from the atom-field covariance matrix \cite{hilico}. To examine the
occurrence of squeezing generated by the system we have represented
in Fig.~\ref{fig:squeezingdelta2} the minimal noise spectra
$S^*(\omega)=\min_{\theta}S_{X_{\theta}}(\omega)$ of different modes
versus the analysis frequency. For a good choice of the interaction
parameters, substantial squeezing can be observed in the $A_{1,2}$
modes, as well as in the ``dark" and ``bright" linearly-polarized
modes \beqr\label{eq:AxAy}
A_x=(A_2-A_1)/\sqrt{2},\hspace{0.3cm}A_y=-i(A_1+A_2)/\sqrt{2}.\eeqr
Since $\langle A_x\rangle=0$, the ``$x$"-polarized component of the
field exiting the cavity is in a squeezed vacuum state, or,
equivalently, in a \textit{polarization-squeezed} state
\cite{korolkova02}, as the quantum noise of an orthogonal component
of the Stokes vector is reduced (Fig.~\ref{fig:SRpolarisation}).\\

The fact that two modes with orthogonal polarization are squeezed at
the output of the cavity is a signature of quantum correlations
existing between orthogonal modes, and therefore of entanglement. A
general method to find out the modes possessing the highest amount
of EPR-type correlations has been outlined in Ref.
\cite{jossejob04}, and experimentally tested in Ref.
\cite{josseprl04}. This method is based on minimizing the quantity
\beqr
\mathcal{E}_{a,b}=\left[\Delta(X_a-X_b)^2+\Delta(Y_a+Y_b)^2\right]/2\eeqr
under unitary transforms on modes $a,b$ (rotations of the
polarization basis), where $X=X_{\theta=0}$ and $Y=X_{\theta=\pi/2}$
are the standard amplitude and phase quadrature operators. The
orthogonal odes $a$ and $b$ are in a non-separable state if
$\mathcal{E}_{a,b}<2$ \cite{duan}.
\begin{figure}
  \includegraphics[width=7.5cm]{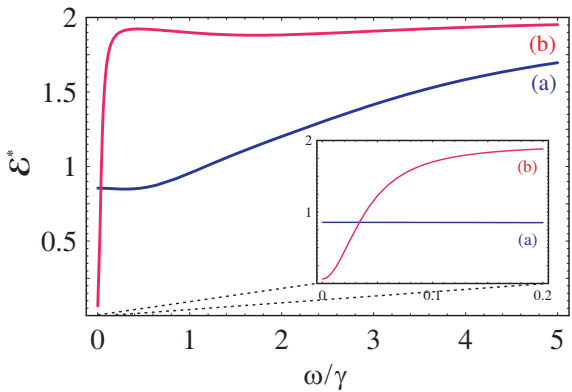}\\
  \caption{Maximal entanglement $\mathcal{E}^*$ that can be obtained
  by adequately mixing the outgoing fields versus sideband frequency $\omega/\gamma$.
  (a) Parameters as in Fig.~\ref{fig:squeezingdelta2}
  (b) $C=1000$, $\kappa=2\gamma$, $\bar{\delta}=0.1$, $\varphi=2$, $I=49$. }\label{fig:intricationdelta2}
\end{figure}
The highest amount of quantum correlations,
$\mathcal{E}^*=\min_{a,b}\mathcal{E}_{a,b}$, that can be produced by
the system is plotted versus frequency in
Fig.~\ref{fig:intricationdelta2} and is of the order of 3 dB for
typical experimental parameters. Numerical simulations show that the
bandwidth over which the squeezing or the entanglement are observed
depends both on the cavity bandwidth $\kappa$ and the atomic
detuning $\delta$. For values of the detuning much smaller than
$\kappa$ and close to the threshold value, the entanglement
bandwidth is given by the CPT width $\delta_s$. Entanglement is thus
larger at zero-frequency and substantially increases with the
cooperativity $C$. Noise reductions of 15 dB can be reached for
$C\sim 1000$ and $\delta\ll\gamma$ (inset in
Fig.~\ref{fig:squeezingdelta2}). However, squeezing - or
entanglement - is produced in a much smaller bandwidth. Note also
that the non-linear Faraday rotation considered here is very
different from the linear self-rotation effect of
Refs.~\cite{josseprl03,hsu} and does not originate from optical
pumping processes. The squeezing is therefore not deteriorated by
excess atomic noise, and much stronger
squeezing can thus be obtained.\\

\begin{figure}
  \includegraphics[width=7.5cm]{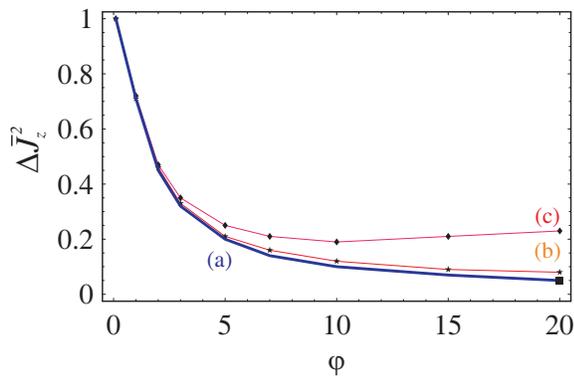}\\
  \caption{$\Delta\bar{J}_z^2$ versus the cavity detuning $\varphi$ for $2\delta=0.01\gamma$:
  (a) Analytical (\ref{eq:Jzapp}), (b) $C=1000$, (c) $C=100$.
  $\alpha$ is chosen to optimize the squeezing for each value of the cavity detuning.}\label{fig:varJz}
\end{figure}

Let us now focus on the atomic fluctuations. For a two-level atom
ensemble polarized along the $x$-axis \textit{spin-squeezing} is
associated to the noise reduction in a spin component in the
$Oyz$-plane below the standard quantum limit given by $\Delta
J_y\;\Delta J_z\geq |\langle J_x\rangle|/2$, and corresponds to the
establishment of quantum correlations between individual spins. The
occurrence of spin-squeezing can be examined by numerically
calculating the minimal variance of spin components in the plane
orthogonal to the mean spin direction. We observe for some values of
the cavity detuning and close to the threshold strong spin-squeezing
in the population difference $J_z$. In order to gain a physical
insight into the process responsible for spin-squeezing we assume a
small detuning so that the spin is mostly polarized along $x$:
$\langle J_x\rangle\simeq -N/2$, and choose the working point close
to the threshold: $\alpha=\delta_s/\delta\succsim 1$ in order to
calculate the fluctuations of $J_z$. Close to the CPT resonance
there are two dominant contributions to the noise of the population
difference $J_z$: first, fluctuations of the optical dipole $P_x$
induced by the \textit{dark} mode $A_x$, secondly, the projection of
the \textit{bright} mode $A_y$ fluctuations due to the Faraday
rotation \beqr\label{eq:Jz} \delta J_z\propto i\Omega\;\delta
P_x-a\;\delta A_y+h.c.+noise,\eeqr with
$a=gN\delta/2\sqrt{2}\Omega$. For small detunings, the population
difference response time - given by (\ref{eq:gammacpt}) - is much
smaller than the optical dipole or the intracavity field evolution
times. Moreover, for a strongly detuned cavity ($\varphi\gg 1$), the
\textit{bright} mode fluctuations $\delta A_y$ will be the sum of
fluctuations due to the \textit{dark} dipole $P_x$ and to the
incident \textit{dark} mode $A_x^{in}$ (damped by a factor
$\varphi$) \beqr \delta A_y\propto ib\;\delta P_x+c\;\delta
A^{in}_x,\eeqr with $b=2\sqrt{2}g/T\varphi$ and
$c=2/\varphi\sqrt{T}$. It is therefore possible for the fluctuations
of the dipole induced by the cavity feedback to compensate the
``natural" fluctuations in (\ref{eq:Jz}) when $\Omega\simeq ab$.
This condition exactly corresponds to the threshold condition
(\ref{eq:seuil}) when $\varphi\gg 1$. After adiabatically
eliminating the optical dipole and the field, it can be shown that
the fluctuations of $J_z$ are proportional to the fluctuations of
the incident \textit{dark} mode amplitude quadrature $X_x^{in}$,
damped by the cavity detuning. The atomic noise spectrum is
Lorentzian-shaped with a width $\gamma_{z}$, given by \beqr
\label{eq:gammacpt}\gamma_{z}
=\delta\sqrt{1+\varphi^2}\left(\alpha-\alpha^{-1}\right).\eeqr
$\gamma_{z}$ clearly vanishes at the threshold ($\alpha=1$) as the
system becomes unstable at the bifurcation point. The normalized
variance in the population difference is then \beqr \Delta
\bar{J}_z^2=\frac{\Delta J_z^2}{N/4}\simeq
\frac{(\alpha\sqrt{1+\varphi^2}-\varphi)^2+1}{(1+\varphi^2)(\alpha^2-1)}.\label{eq:spinsqueezing}\eeqr
Minimizing this quantity with respect to $\alpha$ yields the optimal
spin-squeezing \beqr \label{eq:Jzapp}\Delta \bar{J}_z^*\simeq
\frac{1}{\sqrt{1+\varphi^2}}\hspace{1cm}(C\gg 1,\;\delta\ll
\gamma).\eeqr This simple result expresses that the fluctuations of
the ground state population difference are indeed damped by the
cavity feedback. Because of the Faraday rotation fluctuations of the
\textit{bright} mode $ A_y$ are projected onto the transverse spin
components $J_y$ and $J_z$, the fluctuations of which are either
amplified or reduced depending on the cavity detuning
(Fig.~\ref{fig:SRpolarisation}). A large spin-squeezing can thus be
obtained close to the CPT resonance. The simple expression of
Eq.~(\ref{eq:spinsqueezing}) and the numerical simulation results
are shown in Fig.~\ref{fig:varJz}; an excellent agreement is found
as long as $C\bar{\delta} \gg 1$. When this condition is not
satisfied spontaneous emission noise can no longer be neglected. We
have nevertheless observed that smaller, but still significant,
spin-squeezing can be obtained for larger values of $\bar{\delta}$.

In conclusion, we have presented a cavity scheme based on an atomic
coherence induced non-linear Faraday effect to generate strong
squeezing and entanglement of light fields. Moreover, this scheme
also predicts strong quantum correlations between the light and
atomic variables, resulting in spin-squeezing. Note that the
spin-squeezing mechanism presented here is quite different from
other schemes based upon Faraday rotation \cite{faraday,geremia04},
since it relies on both a strong non-linear interaction as well as a
constructive cavity feedback. We can draw a parallel with the
experiments of Ref.~\cite{geremia04} in which the Faraday
effect-induced fluctuations of a field that has propagated through a
cold atom cloud is fed back to the atoms in order to actively
control the atomic fluctuations. A major difference is that the
feedback is automatically provided by the cavity. Last, the
spin-squeezed state generated could be probed at a later time by
switching back on the fields and performing an adequate measurement
of the fluctuations of the Stokes parameter $S_z$
\cite{dantanpra04}.

We thank Vincent Josse for enlightening discussions. This work was
supported by the COVAQIAL European project No. FP6-511004.

\end{document}